%% file: main.tex
\documentclass[conference]{IEEEtran}
\renewcommand\IEEEkeywordsname{Index Terms}
\usepackage{graphicx}
\usepackage{subcaption}
\usepackage{tikz,xparse}

\usetikzlibrary{patterns}
\usetikzlibrary{dsp,chains}
\usetikzlibrary{matrix}
\usepackage{mathptmx}
\usepackage{verbatim}
\usepackage{calc}
\usepackage{ifthen}
\usepackage{xifthen}
\usepackage{cancel}
\usepackage{bm}
\usepackage{verbatim}
\usepackage{multirow}
\usepackage{cite}
\usepackage[hyphens]{url}
\usepackage[nolist]{acronym} 
\usepackage{pgfplots}
\usetikzlibrary{arrows,shapes,graphs,graphs.standard,quotes,arrows.meta,decorations.markings}
\pgfplotsset{compat=newest}
\usepackage[bookmarks=false]{hyperref}
\usepackage{units}
\usepackage{amsmath, amsbsy, amssymb, latexsym }
\hypersetup{bookmarksdepth=-2}
\usepackage{comment}
\usepackage[utf8]{inputenc}
\usepackage{xcolor}
\usepackage{enumitem}

\usepackage{marginnote}
\tikzset{>=latex}

\input{corporateColours.tex}
\input{nodesAndStyles.tex}

\IEEEoverridecommandlockouts


\begin{document}

\begin{NoHyper}
\title{Learning Quantization in LDPC Decoders}

\author{\IEEEauthorblockN{Marvin Geiselhart\IEEEauthorrefmark{1}, Ahmed Elkelesh\IEEEauthorrefmark{1}, Jannis Clausius\IEEEauthorrefmark{1}, Fei Liang\IEEEauthorrefmark{2}, Wen Xu\IEEEauthorrefmark{3}, Jing Liang\IEEEauthorrefmark{2} and Stephan ten Brink\IEEEauthorrefmark{1}}
	\IEEEauthorblockA{
		\IEEEauthorrefmark{1} Institute of Telecommunications, Pfaffenwaldring 47, University of  Stuttgart, 70569 Stuttgart, Germany 
		\\
		\IEEEauthorrefmark{2} Shanghai Research Center, Huawei Technologies Co., Ltd, Shanghai, China
		\\
		\IEEEauthorrefmark{3} Munich Research Center, Huawei Technologies Duesseldorf GmbH, Riesstr. 25, 80992 Munich, Germany
		\\Email: geiselhart@inue.uni-stuttgart.de\\
	}
}

\maketitle

\begin{acronym}
\acro{ML}{maximum likelihood}
\acro{BP}{belief propagation}
\acro{BPL}{belief propagation list}
\acro{LDPC}{low-density parity-check}
\acro{BER}{bit error-rate}
\acro{SNR}{signal-to-noise-ratio}
\acro{BPSK}{binary phase shift keying}
\acro{AWGN}{additive white Gaussian noise}
\acro{LLR}{log-likelihood ratio}
\acro{MAP}{maximum a posteriori}
\acro{BLER}{block error-rate}
\acro{SCL}{successive cancellation list}
\acro{SC}{successive cancellation}
\acro{BI-DMC}{binary input discrete memoryless channel}
\acro{CRC}{cyclic redundancy check}
\acro{CA-SCL}{CRC-aided successive cancellation list}
\acro{BEC}{binary erasure channel}
\acro{BSC}{binary symmetric channel}
\acro{BCH}{Bose-Chaudhuri-Hocquenghem}
\acro{RM}{Reed--Muller}
\acro{RS}{Reed-Solomon}
\acro{SISO}{soft-in/soft-out}
\acro{3GPP}{3rd Generation Partnership Project }
\acro{eMBB}{enhanced Mobile Broadband}
\acro{CN}{check node}
\acro{VN}{variable node}
\acro{GenAlg}{Genetic Algorithm}
\acro{CSI}{channel state information}
\acro{OSD}{ordered statistic decoding}
\acro{MWPC-BP}{minimum-weight parity-check BP}
\acro{FFG}{Forney-style factor graph}
\acro{MBBP}{multiple-bases belief propagation}
\acro{URLLC}{ultra-reliable low-latency communications}
\acro{DMC}{discrete memoryless channel}
\acro{SGD}{stochastic gradient descent}
\acro{QC}{quasi-cyclic}
\acro{BCE}{binary cross entropy}
\acro{LUT}{look-up table}
\acro{BG}{base graph}
\acro{WS}{weight sharing}
\end{acronym}

\begin{abstract}

Finding optimal message quantization is a key requirement for low complexity \ac{BP} decoding. 
To this end, we propose a floating-point surrogate model that imitates quantization effects as additions of uniform noise, whose amplitudes are trainable variables. 
We verify that the surrogate model closely matches the behavior of a fixed-point implementation and propose a hand-crafted loss function to realize a trade-off between complexity and error-rate performance. 
A deep learning-based method is then applied to optimize the message bitwidths. 
Moreover, we show that parameter sharing can both ensure implementation-friendly solutions and results in faster training convergence than independent parameters.
We provide simulation results for 5G \ac{LDPC} codes and report an error-rate performance within 0.2~dB of floating-point decoding at an average message quantization bitwidth of 3.1~bits. 
In addition, we show that the learned bitwidths also generalize to other code rates and channels.

\end{abstract}
\acresetall



\acresetall

\section{Introduction}

Channel coding is an essential must in any communication system.
The most widely used channel codes nowadays are \ac{LDPC} codes (e.g., in WiMAX, WLAN 802.11n, DVB-S2/T2/C2 and 5G data channel) motivated by the presence of a well-understood \ac{BP} decoder. 
Unfortunately, most of the power dissipation in the receiver side is attributed to the channel decoder operations (i.e., error correction consumes most energy and chip area) \cite{FEC_power}.
Thus, possible improvements might bring large gains in terms of energy efficiency.

\ac{BP} decoding of \ac{LDPC} codes involves passing real-valued infinite precision \ac{LLR} messages between the \acp{VN} and the \acp{CN}.
However, in order to reduce the hardware implementation complexity those messages are typically discretized to a practical finite-precision representation.
On one hand, reducing the message bitwidths degrades the error-rate performance especially at high \acp{SNR} (i.e., error floor region) \cite{SiegelQuantize}.
On the other hand, reducing the bitwidths of the processed quantities inside the decoder leads to a reduced area/memory requirement, possibly increased maximum operating frequency, reduced routing congestion and, thus, high throughput decoder implementations.

In this paper, we focus on optimizing the quantization of the messages involved in the hardware-popular min-sum \ac{LDPC} decoder (i.e., channel \acp{LLR}, \ac{VN}-to-\ac{CN} and \ac{CN}-to-\ac{VN} messages) using machine learning approaches (i.e., \ac{SGD} optimizers).
However, as the quantization is a discontinuous operation whose gradient is zero almost everywhere, we propose the usage of a \emph{surrogate model} for training instead.
It is worth mentioning that this surrogate model approach is also applicable to other applications where quantization optimization is needed.

We investigate the usage of the general concept of the surrogate model in the context of message quantization of an \ac{LDPC} decoder (i.e., learn quantization bitwidths) in order to reduce the overall decoding complexity without sacrificing the error-rate performance. 
This model is used \emph{only} for training (i.e., finding optimal message bitwidths) and is replaced by a regular decoder with ``real'' quantization afterwards.
Note that, here, we focus on decoding \ac{LDPC} codes from the 5G standard, however, it is straightforward to apply the proposed techniques to other (de)coding approaches.

We propose a hand-crafted loss function based on a fixed performance-complexity trade-off parameter $\lambda_c$ (i.e., trade-off between error-rate performance and decoding complexity).
The obtained results use different quantization bitwidths for different messages in the decoding graph which can be interpreted as message importance.
Moreover, we show that weight sharing is possible across different \ac{BP} iterations.

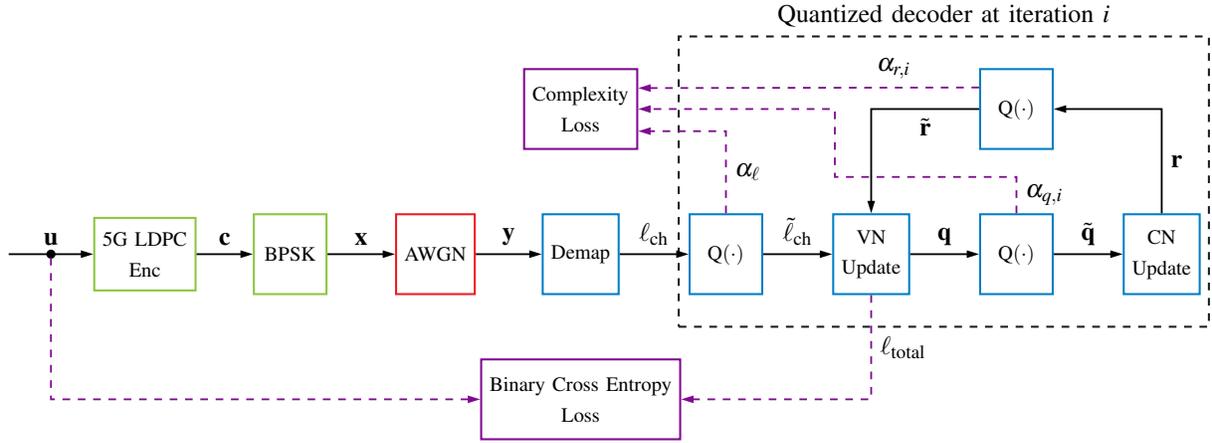
\begin{figure*}[t]
	\centering
	\resizebox{0.9\textwidth}{!}{\input{tikz/system_model_tikz}}
	\caption{\footnotesize{Block diagram of the surrogate model used for learning the quantization bitwidths. 
	A hand-crafted loss function is used consisting of a binary cross entropy (BCE) term and a scaled complexity term, see Eq.~(\ref{eq:hand_crafted_loss}). 
	Note that we use a systematic 5G LDPC code.} }
	\label{fig:surrogate_block}
	\vspace{-0.4cm}
\end{figure*}

\section{Preliminaries}\label{sec:preliminaries}

\subsection{LDPC Codes}

 The key aspect of an \ac{LDPC} code is that the corresponding $(M\times N)$ parity-check matrix $\mathbf{H} = \left [h_{m,l}\right]_{M\times N}$ contains a small number of $1$'s and is therefore sparse. This enables efficient storage of the $\mathbf{H}$-matrix and effective usage of \ac{BP} decoding. The dimensions $M=N-K$ and $N$ define the code rate $R=\nicefrac{K}{N}$, where $K$ is the number of information bits and $N$ is the length of a codeword.

\subsection{Min-Sum Decoding}
The min-sum decoder \cite{fossorier1999minsum} is an iterative message passing decoder that descends from the log-domain version of the \ac{BP} algorithm introduced by Gallager \cite{Gallager}.
The messages, in form of \acp{LLR}, are exchanged over the Tanner graph \cite{Tanner} of the code, which is a bipartite graph  that divides nodes into $N$ \acp{VN} and $M$ \acp{CN}.
We denote messages from \acp{VN} to \acp{CN} by $q_{m,l}$ and messages from \acp{CN} to \acp{VN} by $r_{m,l}$, where the subscript $l~=~{1,...,N}$ indices a certain \ac{VN} and $m={1,...,M}$ a specific \ac{CN}. 
The iterative process of exchanging messages is performed until a stopping condition is met (e.g., in this work decoding stops after $N_\mathrm{iters}$ iterations).
The first $q$ messages are initialized by the channel \acp{LLR} as
\begin{equation*}
    q_{m,l} = \ell_{\mathrm{ch},l},~\forall m.
\end{equation*}
Then, the \acp{CN} are updated and extrinsic $r$ messages are computed as
\begin{equation}\label{eq:CN_update}
    r_{m,l} = \left(\prod_{l'\in V(m) \setminus l} \operatorname{sign}\left( q_{m',l} \right)\right)\cdot \underset{l'\in V(m) \setminus l}{\min} | q_{m',l} |,
\end{equation}
where $V(m)$ is the set of \acp{VN} that is connected to \ac{CN} $m$.
Correspondingly, the update equation of the \acp{VN} is 
\begin{equation*}
    q_{m,l} = \ell_{\mathrm{ch},l} + \sum_{m'\in C(l) \setminus m}r_{m',l},
\end{equation*}
where $C(l)$ is the set of \acp{CN} that are connected to \ac{VN} $l$.
Note that the incoming message $r_{m,l}$ is omitted to calculate the \emph{extrinsic} update of $ q_{m,l}$.
After the final iteration, marginal \acp{LLR} $\ell_{\mathrm{total}}$ are obtained by adding up \textit{all} incoming $r$ messages at each \ac{VN} as
\begin{equation*}
    \ell_{\mathrm{total},l} = \ell_{\mathrm{ch},l} + \sum_{m'\in C(l)}r_{m',l}.
\end{equation*}
A hard decision on $\ell_{\mathrm{total}}$ retrieves the codeword bit estimates of the decoder.
Eq.~(\ref{eq:CN_update}) contains the name-giving approximation by $\min(\cdot)$ function. In contrast, the exact formulation results in a product of costly $\tanh(\cdot)$ functions. 
As the approximated \ac{CN} update results in significantly lower complexity at the price of slightly degraded error-rate performance, the min-sum decoder lays the foundation of state-of-the-art implementations for \ac{LDPC} decoders. Note that the degradation in error-rate performance can be reduced or even nullified by enhancements, such as an attenuation factor or an added offset \cite{chen2005offset_attenuated_ms}. However, as both are straightforward extensions in our quantization framework, we stick to the basic min-sum decoder.

\subsection{Quantized Decoding}

In a hardware-friendly implementation of a message passing \ac{LDPC} decoder (e.g., min-sum decoder) the messages are usually quantized using a uniform quantization with $b$-bits per message representation.
In \cite{chen2005offset_attenuated_ms,StuderLDPC}, $5$-bit and $7$-bit quantizers were used, while in \cite{WehnLDPC} a $4$-bit quantizer was used to implement an unrolled \ac{LDPC} decoder.
Note that typically the message alphabet does not change over the decoding iterations in order to constraint the hardware implementation complexity.
However, in \cite{SiegelQuantize}, a quasi-uniform quantization with an adaptive message bitwidth is used, where the messages involved in later iterations are quantized with a larger bitwidth when compared to earlier iterations. In other words, different message representations are used for different iterations.

A quantized min-sum decoding algorithm for \ac{LDPC} codes is proposed in \cite{AlexiosQuantize1}, where the \ac{VN} update rule is replaced with a \ac{LUT} that is designed using an information-theoretic criterion (e.g., using density evolution).
Optimizing the \ac{LUT}-based \ac{VN} update can accommodate lower bitwidths for all messages involved in the decoding process without sacrificing the error-rate performance (e.g., 3-bit quantizers for internal messages and 4-bit quantizers for the channel \acp{LLR}).

Another method of designing finite-precision decoders is the \emph{information bottleneck} algorithm.
Again, the aim is to design lower complexity decoding algorithms with error-rate performance close to high-precision decoders.
Information bottleneck can be used to design \ac{LUT}-based node update computations.
For more details we refer the interested reader to \cite{IB_Bauch} and the references therein.

\section{Learning Quantization}

\subsection{Quantizer Implementation}
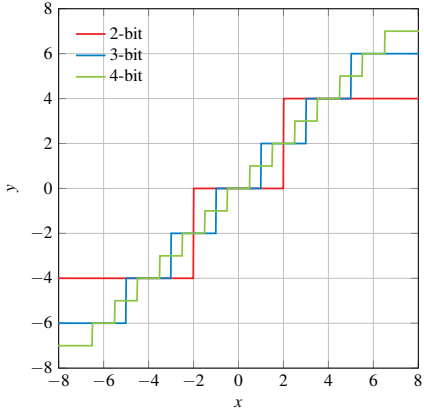
\begin{figure}
    \centering
    \resizebox{0.65\columnwidth}{!}{\input{tikz/quantization_curves}}
    \caption{\footnotesize Quantization curves for the proposed ``compatible'' quantization scheme, $L_\mathrm{limit}=8$.}
    \label{fig:quantcurve}
    \vspace{-.5cm}
\end{figure}

For maximum efficiency, we allow each value in the decoding algorithm to take a different bitwidth. This way, the memory, wiring and circuit complexity maybe minimized. However, arithmetic operations may require their operands to be the same bitwidth. Therefore, we desire that conversion from a lower bitwidth to a higher bitwidth can be done without any overhead. 
For this, we first formulate the constraint that the set of quantization levels of lower bitwidths is always a subset of the set of quantization levels of higher bitwidths. If $\mathcal{Q}_b$ denotes the set of quantization levels for a given bitwidth $b$, this constraint can be expressed as 
\begin{equation*}
	\mathcal{Q}_b \subseteq \mathcal{Q}_{b+1}.
\end{equation*}
Moreover, $0$ should be a valid quantization level (for all bitwidths) and the quantizer should be symmetric. Thus, there are $2^b-1$ valid quantization levels and the quantization step size is defined as
\begin{equation*}
	\alpha(b) = \frac{L_\mathrm{limit}}{2^{b-1}},
\end{equation*}
where $L_\mathrm{limit}$ is an overall scaling parameter that defines the dynamic range of the quantizer
\begin{equation*}
	L_{\mathrm{limit}} = \lim_{b\to \infty} \max\{\mathcal{Q}_b\}.
\end{equation*}
Note that the maximum representable value is dependent on $b$ and given by
\begin{equation*}
	L_{\mathrm{clip},b} = \max\{\mathcal{Q}_b\} = L_\mathrm{limit} - \alpha(b).
\end{equation*}
Fig.~\ref{fig:quantcurve} shows the quantization curves for the proposed scheme for $L_\mathrm{limit}=8$ and $b\in\{2,3,4\}$ bits. Note that a two's complement labeling works naturally with this quantizers, as going to a larger bitwidth corresponds to appending zeros.

\subsection{Surrogate Model}

Fig. \ref{fig:surrogate_block} shows the block diagram of the considered communications model, consisting of \ac{LDPC} encoding, \ac{BPSK} mapping, \ac{AWGN} channel, demapping and min-sum decoding. The goal is finding suitable bitwidths for a fixed-point implementation of the decoder. However, in our proposed approach, we use a floating-point version with a quantizer after each operation, as indicated by blocks labeled $Q(\cdot)$. In particular, we quantize the channel \ac{LLR} $\ell_\mathrm{ch}$, the VN-to-CN messages $q$ and the CN-to-VN messages $r$. The corresponding quantization steps are the independent trainable parameters $\alpha_{q,i,e}$, $\alpha_{r,i,e}$ and $\alpha_{\ell,l}$, where $i$ denotes the iteration and $e$ denotes the edge of the Tanner graph $(m,l)$ with $h_{m,l}=1$.

To optimize the quantization steps (and, thus, the bitwidths) of the quantizers, we propose to use a surrogate model, where each quantizer is replaced by the addition of a random variable $n$. This random variable models the quantization error (or quantization noise). We use the common assumption that the quantization noise for a uniform quantizer is uniformly distributed and independent of the input signal \cite{GrayNeuhoffQuant}. For a quantization with quantization step $\alpha$ that rounds to the nearest quantization level, the quantization error is therefore distributed according to $\mathrm{U}(-0.5\alpha,0.5\alpha $). Equivalently, we can draw $n$ from $\mathrm{U}(-0.5,0.5$) and scale it by $\alpha$. 
Lastly, we account for the limited number of quantization steps by clipping the result of the addition at $\pm L_\mathrm{clip}$. A block diagram of a single surrogate quantizer is shown in Fig.~\ref{fig:surrogate_section}. Each quantizer in the surrogate model has its own trainable parameter $\alpha$ that can be optimized using gradient-based methods. For this, we need a suitable differentiable loss function with respect to the trainable parameters $\alpha$.

\subsection{Loss Function}

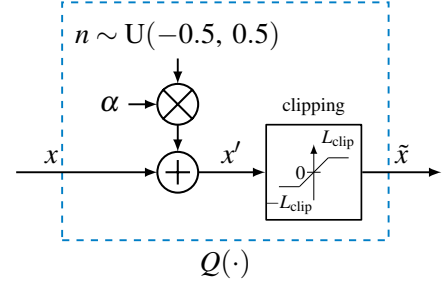
\begin{figure}[t]
	\centering
	\resizebox{0.7\columnwidth}{!}{\input{tikz/surrogate_model_tikz}}
	\caption{\footnotesize{A single quantizer $\operatorname{Q}(\cdot)$ of the surrogate model used for training the quantization bitwidths.}}
	\label{fig:surrogate_section}
	\vspace{-0.5cm}
\end{figure}

A suitable loss function to optimize the error-rate performance using bit-metric decoding is the \ac{BCE} loss \cite{cammerer20tcom}. The reason is that the minimization of the loss is equivalent to maximizing the bit-wise mutual information.
As $\mathbf{\ell}_{\mathrm{total},l}$ are the logits of the message estimate, we apply the sigmoid function in the loss computation. We assume that the code is systematic and, thus, the first $K$ indices of $\mathbf{\ell}_{\mathrm{total}}$ correspond to the information vector $\mathbf{u}$. As a result, the total loss function is given by
\begin{equation}\label{eq:hand_crafted_loss}
	\mathcal{L} = \sum_{l=1}^{K}\operatorname{BCE}(\operatorname{sigmoid}(\ell_{\mathrm{total},l}), u_l) + \lambda_\mathrm{c}\cdot C,
\end{equation}
where $C$ models the (bitwidth-based) complexity of the decoding and $\lambda_c$ is a fixed hyper parameter specifying the trade-off between error-rate performance and complexity. In practice, Eq. (\ref{eq:hand_crafted_loss}) is estimated using batches of $B$ transmissions of random codewords over the simulated channel. $C$ is defined as the mean of the corresponding quantizer bitwidths
\begin{align*}\label{eq:complexity}
    C =& \frac{1}{2 N_\mathrm{iters} \cdot N_\mathrm{msg} + N \cdot N_\mathrm{iters} }  \Bigg. \Bigg[
    N_\mathrm{iters} \sum_{l=1}^{N}  \Bigg( \log_2\left(\frac{L_\mathrm{limit}}{\alpha_{\ell,l}}\right)+1 \Bigg) \nonumber\\
    &+\sum_{i=1}^{N_\mathrm{iters}} \sum_{e=1}^{N_\mathrm{msg}} \Bigg( \Bigg.  \log_2\left(\frac{L_\mathrm{limit}}{\alpha_{q,i,e}}\right) +1  +\log_2\left(\frac{L_\mathrm{limit}}{\alpha_{r,i,e}}\right) +1 \Bigg) \Bigg. \Bigg] \Bigg..
\end{align*}
As the channel \acp{LLR} $\ell_\mathrm{ch}$ are required in every iteration, they are counted $N_\mathrm{iters}$ times in the complexity formula.

The surrogate model is then trained (i.e., the parameters $\alpha_{q,i,e}$, $\alpha_{r,i,e}$ and $\alpha_{\ell,l}$) using a \ac{SGD}-based optimizer, e.g., Adam~\cite{adam}, with batches of $B$ transmissions per epoch.

\subsection{Model Conversion}
To convert the trained surrogate model to a fixed-point implementation (i.e., real quantizers), the real-valued parameters $\alpha_{q,i,e}$, $\alpha_{r,i,e}$ and $\alpha_{\ell,l}$ are converted to the corresponding bitwidth according to
\begin{equation*}
    b(\alpha) = \operatorname{round}\left( \log_2\left(\frac{L_\mathrm{limit}}{\alpha}\right) +1 \right),
\end{equation*}
where $\operatorname{round}(x)$ returns the closest integer to $x$.

\section{Results}
We evaluate the surrogate model-based quantization bitwidth training using the 5G \ac{LDPC} code \cite{5GLDPC,Richardson5G} with parameters $N=264$ and $K=132$, which corresponds to the second \ac{BG} using lifting size $Z=22$. We fix the number of iterations to $N_\mathrm{iters}=10$.
\subsection{Surrogate Model Performance}
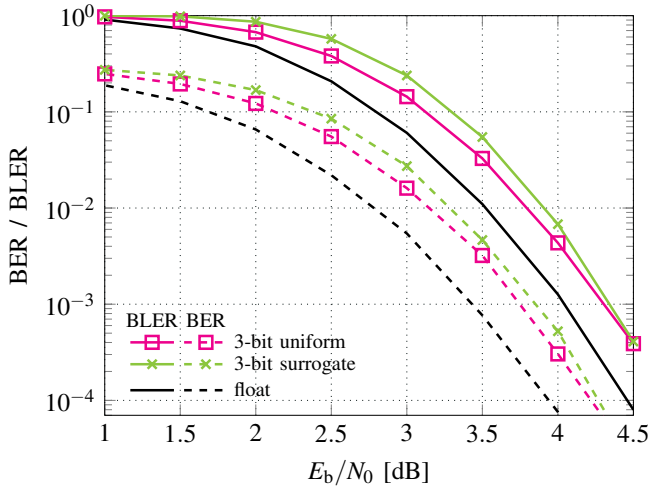
\begin{figure} [t]
	\centering
	\resizebox{\columnwidth}{!}{\input{tikz/uniform_vs_surrogate}}
	\caption{\footnotesize Error-rate performance comparison of unquantized, quantized and surrogate min-sum decoding for the (264,132) 5G LDPC code.}
	\label{fig:uvss}
\end{figure}

First, we assess how close the surrogate model matches a real fixed-point quantization. In Fig. \ref{fig:uvss}, we compare the error-rate performance of a 3-bit fixed-point model with the (untrained) surrogate model with an equivalent bitwidth of 3-bit. As a baseline, we also plot the performance of a floating-point implementation. We see that there is a gap of approximately 0.3 dB of the fixed-point implementation to the floating-point baseline. Moreover, the surrogate model performance is slightly worse than that of the fixed-point implementation. We can explain that by the mismatch of the distribution of the quantization error, which is not quite uniform in the actual fixed-point scenario. Hence, the surrogate model's quantization error has a larger variance than the real model.

\subsection{Trained Model}
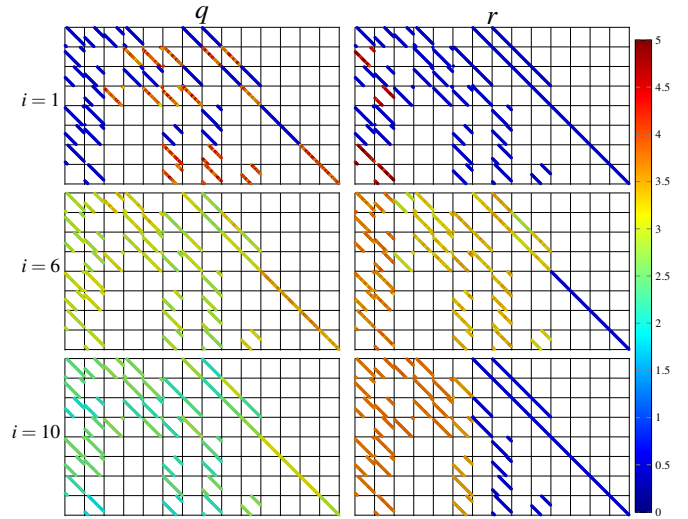
\begin{figure} [t]
    \vspace{0.2cm} 
	\centering
	\resizebox{\columnwidth}{!}{\input{tikz/h_matrices}}
	\caption{\footnotesize Learned quantization bitwidths (color-coded) for the $q$ messages (\emph{left}) and $r$ messages (\emph{right}) in the first, sixth and tenth iteration for the $(264,132)$ 5G LDPC code. No weight sharing is applied, grid size is $Z=22$. Note that dark blue points correspond to a bitwidth of 0, implying that this message is zero-valued.}
	\label{fig:bitwidths}
	\vspace{-0.4cm}
\end{figure}
For training, we set the hyper parameter $\lambda_c=0.15$, fix $L_\mathrm{limit}=8$ and initialize all trainable parameters to $\alpha=1$, corresponding to 4-bit quantization. The system is trained in an \ac{AWGN} scenario at an \ac{SNR} of $E_\mathrm{b}/N_0=2.5\text{ dB}$ using Adam \cite{adam} with learning rate $\gamma=2.5\cdot10^{-3}$ and a batch size $B=2048$ for 3000 epochs. The resulting bitwidths (not rounded) for $q$ and $r$ are depicted in Fig. \ref{fig:bitwidths} for iterations $i=1,6,10$, respectively. The values are arranged according to the parity-check matrix $\mathbf{H}$ of the code. We observe that in the first iteration, only very few checks are important; the other checks do not require any accuracy (i.e., the values are just 0). This can be explained by the punctured message bits of the 5G \ac{LDPC} code. The first $2Z = 44$ \acp{VN} are initialized with \acp{LLR} $=0$, therefore, all checks involving two of these punctured \acp{VN} only output zero-valued messages. Only the $q$ and $r$ values needed to reconstruct the punctured message bits are active. 
Similarly, the $r$ messages of the extension parity-bits are quantized with a low resolution. This makes sense, as the extension \acp{VN} are degree-1 and therefore, the values are not used by the \ac{VN} update. The $q$ messages outgoing of the extension parity-bits are generally more important than other messages and are quantized with a higher resolution. Finally, in the last iteration, only the $r$ messages corresponding to the $K$ message bits are quantized with a reasonable bitwidth. This is easily explained by the fact that those are the only bits used in the \ac{BCE} loss calculation. The learned bitwidths for the channel \acp{LLR} remained approximately at 4, resulting in an overall average complexity of 3.1 bits per message.

\subsection{Weight Sharing}
\begin{figure} [t]
	\centering
	\resizebox{\columnwidth}{!}{\input{tikz/loss}}
	\caption{\footnotesize Loss vs. training epoch for different weight sharing methods, using the (264,132) 5G LDPC code.}
	\label{fig:loss}
\end{figure}
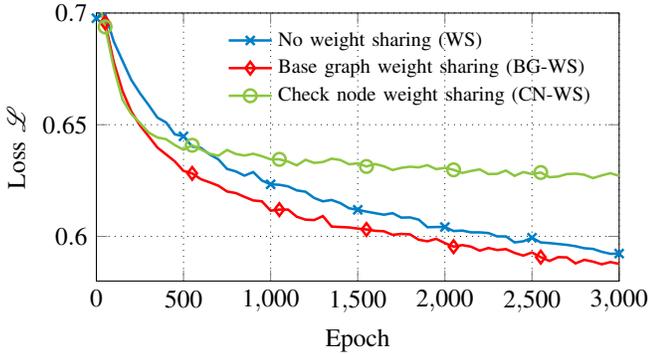
\begin{table}[t]
	\centering
	\begin{tabular}{l|ccc}
		Method & No WS & BG-WS & CN-WS \\
		\hline
		\# of parameters & 19228 & 874 & 161 \\
		Avg. bitwidth, surrogate & 3.289 & 3.304 & 3.486 \\
		Avg. bitwidth, converted & 3.103 & 3.126 & 3.452 \\
	\end{tabular}
	\caption{\footnotesize Resulting complexity of different weight sharing methods for the (264,132) 5G LDPC code.}
	\label{tab:ws_comp-1}
\end{table}
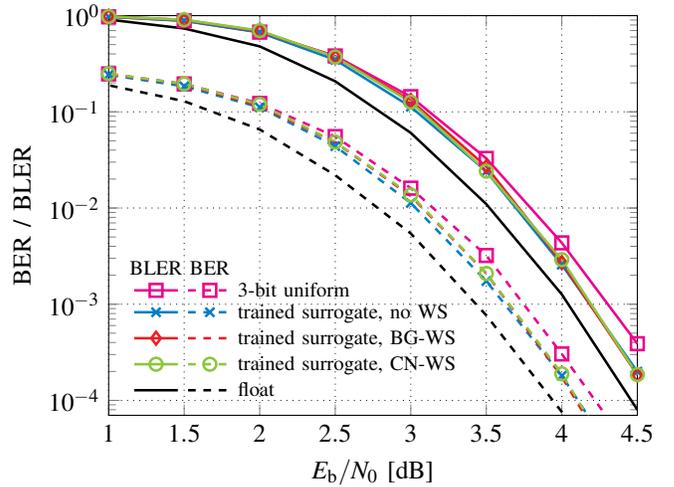
\begin{figure} [t]
	\centering
	\resizebox{\columnwidth}{!}{\input{tikz/results_ws_er}}
	\caption{\footnotesize Error-rate performance comparison of the trained surrogate decoder for different weight sharing methods, using the (264,132) 5G LDPC code.}
	\label{fig:er_ws}
\end{figure}

\begin{figure} [t]
	\centering
	\resizebox{\columnwidth}{!}{\input{tikz/results_ws_er_conv}}
	\caption{\footnotesize Error-rate performance comparison of the converted decoder for different weight sharing methods, using the (264,132) 5G LDPC code.}
	\label{fig:er_ws_conv}
	\vspace{-0.3cm}
\end{figure}
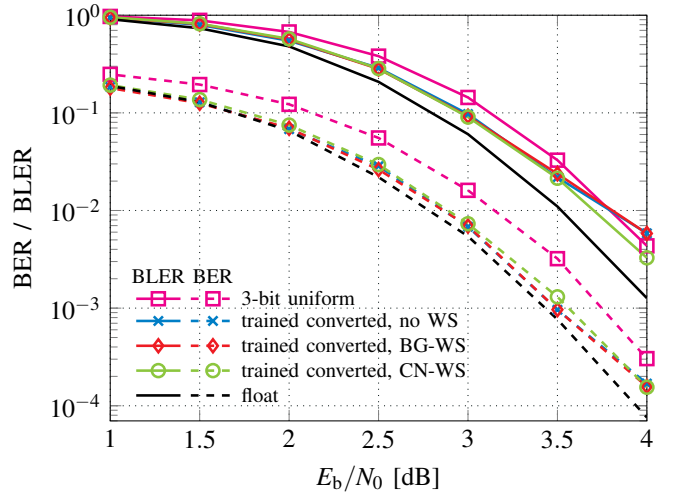

One can observe from Fig.~\ref{fig:bitwidths} that messages corresponding to the same \ac{BG} edge should be quantized with a similar bitwidth. Therefore it is reasonable to share a single trainable parameter for $Z$ edges, reducing the training complexity. Another possible \ac{WS} approach is to share the same bitwidth across each \ac{BG}-\ac{CN} (i.e., groups of $Z$ rows in the parity-check matrix). The number of parameters and their resulting complexity (after training and conversion) is listed in Tab.~\ref{tab:ws_comp-1}. Fig.~\ref{fig:loss} shows the development of the loss value $\mathcal{L}$ over the training epochs. We can see that \ac{WS} does not only reduce the number of trainable parameters, it also speeds up the model convergence. While \ac{CN}-based \ac{WS} converges similarly fast, it cannot reach loss values as low as the other methods. This is expected by the previously observed diversity in the message bitwidths of the inbound and outbound messages to one \ac{CN}. Consequently, training cannot achieve complexity values as low as in the other \ac{WS} approaches. The error-rate performance of the surrogate and converted decoders are depicted in Fig.~\ref{fig:er_ws} and Fig.~\ref{fig:er_ws_conv}, respectively. We do not observe any difference in error-rate performance in case of the surrogate models for different \ac{WS} methods. However, the converted model with \ac{CN}-\ac{WS} seems to achieve a lower error floor than the other methods.

\subsection{Generalization}
\begin{figure}[t]
	\centering
	\resizebox{\columnwidth}{!}{\input{tikz/results_rayleigh}}
	\vspace{-0.2cm}
	\caption{\footnotesize Error-rate performance of converted decoder model trained for AWGN and Rayleigh fading channel, evaluated on a Rayleigh fading channel with the (264,132) 5G LDPC code. Both trained models have an average bitwidth of $3.1$ bits.}
	\label{fig:dec_surrogate_rayleigh}
	\vspace{-0.3cm}
\end{figure}
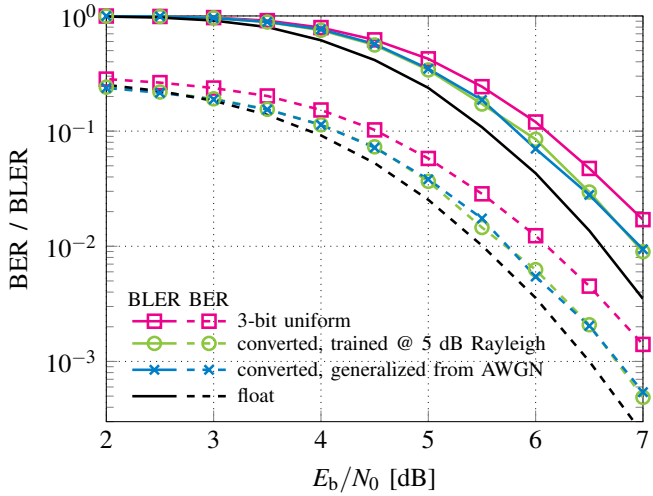

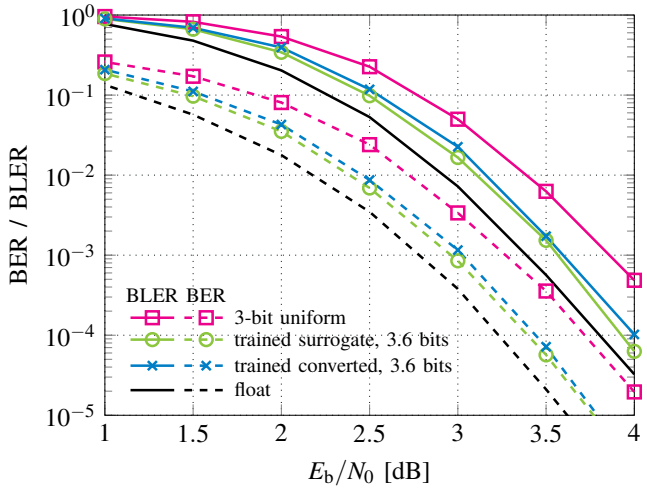
\begin{figure}[t]
	\centering
	\resizebox{\columnwidth}{!}{\input{tikz/results_1_4} \vspace{-0.2cm}}
	\caption{\footnotesize Error-rate performance of the surrogate and converted trained model for the (528,132) 5G LDPC code with $R = 1/4$.}
	\label{fig:quc_nov_res_1_4}
	\vspace{-0.5cm}
\end{figure}

\begin{figure}[t]
	\centering
	\resizebox{\columnwidth}{!}{\input{tikz/results_2_3}}
	\vspace{-0.2cm}
	\caption{\footnotesize Error-rate performance of the surrogate and converted trained model for the (198,132) 5G LDPC code with $R' = 2/3$ as well as the generalized converted model from $R = 1/4$. Both trained models have an average bitwidth of $3.95$ bits.}
	\label{fig:quc_nov_res_2_3}
	\vspace{-0.3cm}
\end{figure}
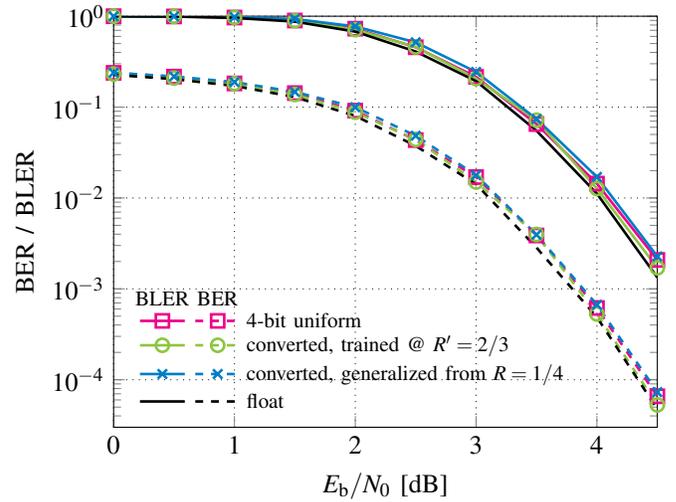

Finally, we evaluate how the proposed method can generalize to other channels or rates. For this, we train a model for one scenario and evaluate the learned parameters in a different scenario. Fig.~\ref{fig:dec_surrogate_rayleigh} shows the performance of the \ac{AWGN} trained model in an ergodic Rayleigh fading channel scenario with perfect channel state information. As we can see, the generalized model has almost the same performance as a model that was trained in this environment (same complexity of $3.1$ bits, using $\lambda_c=0.15$ at $E_\mathrm{b}/N_0 = 5 \text{ dB}$).

Next, we evaluate how the model can generalize to other code rates. 
For this we train the surrogate model at a low code rate of $R=1/4$, corresponding to $K=132$ and $N=528$, using $\lambda_c=0.003$. 
This performance is shown in Fig.~\ref{fig:quc_nov_res_1_4}; the corresponding complexity is $3.6$ bits. 
Then, we take the bitwidths from this model and evaluate it for a rate of $R'=2/3$, i.e., $N'=198$. 
This is possible, as in the 5G standard, all codes with the same code dimension use the same parity-check matrix, and merely differ in the number of transmitted/punctured extension parity-bits.
Fig.~\ref{fig:quc_nov_res_2_3} shows the results where again it can be seen that the model performs very similarly to the one trained at this rate (same complexity of $3.95$ bits, using $\lambda_c=0.011$ at $E_\mathrm{b}/N_0 = 3 \text{ dB}$). 
Note that the higher rate code has fewer of the coarsely quantized extension checks, increasing its average complexity.

\section{Conclusion and Outlook}\label{sec:conc}
We propose a novel method to optimize the bitwidths of a fixed-point min-sum \ac{LDPC} decoder. 
The method is based on a surrogate model where quantization has been replaced by the addition of noise, such that the model remains differentiable and gradient-based optimization can be applied. 

Weight sharing across base graph edges reduces the training complexity and improves the convergence speed. 
Furthermore, as all parameters are on base graph level, it allows for easy generalization to other blocklengths and rates via the standard rate-matching schemes.

The proposed methodology can be straightforwardly extended to attenuated or offset min-sum decoding \cite{chen2005offset_attenuated_ms}. 
Moreover, many other parts of a communications system, where fixed-point implementations are used, can be optimized using the surrogate approach.

\bibliographystyle{IEEEtran}
\bibliography{references}
\end{NoHyper}
\end{document}

%% file: corporateColours.tex
\definecolor{mittelblau}{RGB}{0, 126, 198}
\definecolor{violettblau}{cmyk}{0.9, 0.6, 0, 0}
\definecolor{rot}{RGB}{238, 28 35}
\definecolor{apfelgruen}{RGB}{140, 198, 62}
\definecolor{gelb}{RGB}{255, 229, 0}
\definecolor{orange}{RGB}{244, 111, 33}
\definecolor{pink}{RGB}{237, 0, 140}
\definecolor{lila}{RGB}{128, 10, 145}
\definecolor{hellgrau}{RGB}{224, 224, 224}
\definecolor{mittelgrau}{RGB}{128, 128, 128}
\definecolor{dunkelgrau}{RGB}{80,80,80}
\definecolor{anthrazit}{RGB}{19, 31, 31}
\definecolor{darkgreen}{RGB}{34,139,34}
\definecolor{aqua}{RGB}{0, 255, 255}

%% file: nodesAndStyles.tex
\tikzset{
       vnd/.style={
        shape=circle,
        fill=black,
        draw,
        inner sep=0pt,
        minimum size=0.2cm},
        cnd/.style={
        shape=rectangle,
        fill=white,
        draw,
        minimum width=0.05mm,
        minimum height = 0.05mm}, 
         vndR/.style={
        shape=circle,
        fill=red,
        draw,
        inner sep=0pt,
        minimum size=0.2cm},
        cndR/.style={
        shape=rectangle,
        fill=white,
        draw=red,
        minimum width=0.05mm,
        minimum height = 0.05mm}
}

%% file: tikz/system_model_tikz.tex
\begin{tikzpicture}

\tikzstyle{dots}=[line width=1pt , line cap=round,dash pattern=on 0pt off 4pt,black]

\tikzstyle{arrow} = [thick,->, rounded corners=5pt]

\tikzstyle{loosely dotted}=          [dash pattern=on \pgflinewidth off 8pt]

\tikzstyle{NN} =[rectangle, rounded corners,   align=flush center,minimum width=3cm,minimum height=1.0cm,text centered,
	draw=red, fill=red!0 , thick , label={[red]:NN}];

\tikzstyle{box1} =[rectangle, dashed,  align=flush center,minimum width=4.5cm, minimum height=4.5cm,text centered,
	draw=blue, fill=red!0 , thick , label={[black]:}];	
\tikzstyle{boxText} =[rectangle,  align=flush center,minimum width=15.74776cm,
	minimum height=15.74776cm,text centered,
	draw=black!50, fill=red!0 , thick ];
\tikzstyle{interleaver} =[rectangle, rounded corners,   align=flush center,minimum width=1.0cm, minimum height=1.0cm,text centered,
	draw=black, fill=red!0 , thick, text=black];
\tikzstyle{plus} =[circle,draw=black, fill=white, inner sep=2pt,minimum size=5pt];
	
\tikzstyle{box} =[rectangle,   align=flush center,minimum width=1.0cm, minimum height=1.1cm,text centered,
	draw=black, fill=red!0 , thick, text=black];
	
\def\distance{2}

\node [box1, draw=anthrazit, minimum width=7.3cm, minimum height=4.cm, label={[black]:Quantized decoder at iteration $i$}] at (6.5*\distance,1) {};
	
\node [] (start) at (0* \distance,0) {};

\node [box, draw= apfelgruen,align=center,minimum height=1.0cm] (enc) at (1* \distance,0) {\footnotesize 5G LDPC \\ \footnotesize Enc};

\node [box, draw= apfelgruen] (map) at (2* \distance,0) {\footnotesize BPSK};

\node [box, draw= rot] (awgn) at (3* \distance,0) {\footnotesize AWGN};

\node [box, draw=mittelblau] (demap) at (4* \distance,0) {\footnotesize Demap};

\node [box, draw=mittelblau] (q1) at (5* \distance,0) {\footnotesize $\operatorname{Q}(\cdot)$};

\node [box, draw=mittelblau] (VN) at (6* \distance,0) {\footnotesize VN \\ \footnotesize Update};

\node [box, draw=mittelblau] (q2) at (7.* \distance,0* \distance) {\footnotesize $\operatorname{Q}(\cdot)$};
\node [box, draw=mittelblau] (q3) at (7.* \distance,1* \distance) {\footnotesize $\operatorname{Q}(\cdot)$};

\node [box, draw=mittelblau] (CN) at (8* \distance,0* \distance) {\footnotesize CN \\ \footnotesize Update};

\node [box, draw=lila] (BCE) at (4* \distance,-1* \distance) {\footnotesize Binary Cross Entropy \\ \footnotesize Loss};

\node [box, draw=lila] (C) at (4* \distance,1* \distance) {\footnotesize Complexity \\ \footnotesize Loss};

\node [] (end) at (6* \distance,-2) {};

\draw [dspconn] (start) -- node [dspnodefull] (inter1) {} node[above]{$\mathbf{u}$}(enc);
\draw [dspconn] (enc) -- node[above]{$\mathbf{c}$}(map);
\draw [dspconn] (map) -- node[above]{$\mathbf{x}$}(awgn);
\draw [dspconn] (awgn) -- node[above]{$\mathbf{y}$}(demap);
\draw [dspconn] (demap) -- node[above]{$\mathbf{\ell}_{\mathrm{ch}}$}(q1);
\draw [dspconn] (q1) -- node[above]{$\tilde{\mathbf{\ell}}_{\mathrm{ch}}$}(VN);
\draw [dspconn] (VN) -- node[above]{$\mathbf{q}$}(q2);
\draw [dspconn] (q2) -- node[above]{$\tilde{\mathbf{q}}$}(CN);
\draw [dspconn] (CN) |- node[near start, right]{$\mathbf{r}$}(q3);
\draw [dspconn] (q3) -| node[near start,below ]{$\tilde{\mathbf{r}}$}(VN);

\draw [dspconn, dashed, draw=lila] (VN)  |- node[near start,right]{$\mathbf{\ell}_{\mathrm{total}}$} (BCE);

\draw [dspconn, dashed, draw=lila] (inter1) |- node[above]{}(BCE);

\draw [dspconn, dashed, draw=lila] (q3.150) --  node[near start, above]{$\mathbf{\alpha}_{r,i}$}(C.20);
\draw [dspconn, dashed, draw=lila] (q1.north) |- node[near start, right]{$\mathbf{\alpha}_{\ell}$}(C.base east);

\draw [dspconn, dashed, draw=lila] (q2.north) -- node[right]{$\mathbf{\alpha}_{q,i}$} ($(0,0.5)+(q2.north)$) -- ($(-2.5,0.5)+(q2.north)$) -- ($(-2.5,2)+(q2)$) -- node[above]{}(C);
	
\end{tikzpicture}

%% file: tikz/quantization_curves.tex
\begin{tikzpicture}
\begin{axis}[
        width=\linewidth,
        height=\linewidth,
        xmajorgrids,
        xminorgrids=true,
        yminorticks=true,
        ymajorgrids,
        yminorgrids,
		xmin=-8,
		xmax=8,
		ymin=-8,
		ymax=8,
        ymode=normal,
        legend style={legend cell align=left,align=left,fill opacity=0, text opacity=1, draw=none,at={(0.28,0.8)}},
        legend pos=north west,
        ylabel={$y$},
        xlabel={$x$},
    ]

\addplot+ [rot,line width=1pt , mark options={opacity=0.0,fill=rot, solid}] 
table[x=x,y=y ,col sep=comma]{./tikz/quantizer/quantizer_with_bitwidth_2.txt};
\addlegendentry{2-bit };

\addplot+ [mittelblau,line width=1pt , mark options={opacity=0.0,fill=rot, solid}] 
table[x=x,y=y ,col sep=comma]{./tikz/quantizer/quantizer_with_bitwidth_3.txt};
\addlegendentry{3-bit };

\addplot+ [apfelgruen,line width=1pt , mark options={opacity=0.0,fill=rot, solid}] 
table[x=x,y=y ,col sep=comma]{./tikz/quantizer/quantizer_with_bitwidth_4.txt};
\addlegendentry{4-bit };
\end{axis}

\end{tikzpicture}

%% file: tikz/surrogate_model_tikz.tex
\begin{tikzpicture}

\tikzstyle{dots}=[line width=1pt , line cap=round,dash pattern=on 0pt off 7pt,orange]

\tikzset{font={\fontfamily{ptm}\fontsize{12pt}{14}\selectfont}}

\tikzstyle{arrow} = [thick,->]

\tikzstyle{loosely dotted}=          [dash pattern=on \pgflinewidth off 8pt]

\tikzstyle{NN} =[rectangle, rounded corners,   align=flush center,minimum width=3cm,minimum height=1.7cm,text centered,
	draw=red, fill=red!0 , thick , label={[red]:NN}];
	
\tikzstyle{box} =[rectangle, rounded corners,   align=flush center,minimum width=1.7cm, minimum height=1.7cm,text centered,
	draw=black, fill=red!0 , thick];
	
\tikzstyle{boxText} =[rectangle,  align=flush center,minimum width=15.74776cm,
	minimum height=15.74776cm,text centered,
	draw=black!50, fill=red!0 , thick ];
	
\tikzstyle{plus} =[circle,draw=black, fill=white, inner sep=1,minimum size=30pt, font=\huge];

\tikzstyle{box1} =[rectangle, dashed,  align=flush center,minimum width=4.5cm, minimum height=5.0cm,text centered,
	draw=blue, fill=red!0 , thick , label={[black]:}];	
\node [box1, draw=mittelblau, minimum width=4.8cm, minimum height=3.5cm, label={below:\color{black}$Q(\cdot)$ }] at (2.7,2.75) {};

	\node [dspadder, minimum size=0.6cm] (Demapper) at (2,2) {};
	\node [dspsquare, minimum size=1.4cm, label=above:\footnotesize clipping] (conv_1) at (4,2) {};
	
	\draw ($ (conv_1.south west) + (0.5,0.5) $) -- ($ (conv_1.north east) + (-0.5,-0.5) $);
	
	\draw ($ (conv_1.south west) + (0.5,0.5) $) -- node[below]{\footnotesize $-L_{\mathrm{clip}}$} ($ (conv_1.south west) + (0.2,0.5) $);
	
	\draw ($ (conv_1.north east) + (-0.5,-0.5) $) -- node[above]{\footnotesize $L_{\mathrm{clip}}$}($ (conv_1.north east) + (-0.2,-0.5) $);
	
	\draw[->] ($ (conv_1.south) + (0.0,0.3) $) -- ($ (conv_1.north) + (+0.0,-0.3) $);
	
	\draw ($ (conv_1) + (-0.1,0.0) $) -- node[left]{\footnotesize$0$}($ (conv_1) + (+0.1,-0.0) $);

	\node [dspmixer, minimum size=0.6cm] (conv_2) at (2,3) {};
	\node [] (end) at (6,2) {};

	\node [label=below :] (h) at (1,3) {$\alpha $};
	\node [label=above:$n \sim \mathrm{U} (-0.5\text{,~} 0.5)$] (alpha) at (2,3.55) {};
	\node [label=] (start) at (-0.5,2) {};

	\draw [arrow] (start) -- node[ near start,above] {$x$} (Demapper)  ;
	\draw [arrow] (Demapper) -- node[ above] {$x'$}     (conv_1)  ;
	\draw [arrow] (conv_1) -- node[ above] {$\Tilde{x}$}     (end)  ;
	
	\draw [arrow] (h) -- node[ near start,above] {}   (conv_2)  ;
	\draw [arrow] (alpha.north) -- node[ above] {}     (conv_2)  ;
	\draw [arrow] (conv_2) -- node[ above] {}     (Demapper)  ;

\end{tikzpicture}

%% file: tikz/uniform_vs_surrogate.tex
\begin{tikzpicture}
\begin{axis}[
width=\linewidth,
height=.8\linewidth,
grid style={dotted,anthrazit},
xmajorgrids,
yminorticks=true,
ymajorgrids,
legend columns=1,
legend pos=south west,   
legend cell align={left},
legend style={fill,fill opacity=0, text opacity=1, draw=none},
xlabel={$E_\mathrm{b}/N_0$ [dB]},
ylabel={BER / BLER},
legend image post style={mark indices={}},	
ymode=log,
mark size=1.5pt,
xmin=1,
xmax=4.5,
ymin=7e-5,
ymax=1,
]

\addplot[color=magenta,line width = 1pt, solid,mark=square,mark size=2.5pt, mark options={solid}]
table[col sep=comma]{
0.00, 9.993e-01
0.50, 9.973e-01
1.00, 9.719e-01
1.50, 8.860e-01
2.00, 6.741e-01
2.50, 3.811e-01
3.00, 1.438e-01
3.50, 3.278e-02
4.00, 4.344e-03
4.50, 3.888e-04
};
\label{plot:3bit_bler}

\addplot[color=apfelgruen,line width = 1pt, solid,mark=x,mark size=2.5pt, mark options={solid}]
table[col sep=comma]{
0.00, 1.000e+00
0.50, 9.995e-01
1.00, 9.978e-01
1.50, 9.846e-01
2.00, 8.645e-01
2.50, 5.740e-01
3.00, 2.388e-01
3.50, 5.477e-02
4.00, 6.795e-03
4.50, 4.096e-04
};
\label{plot:surr_bler}

\addplot[color=black,line width = 1pt, solid,mark size=2.5pt, mark options={solid}]
table[col sep=comma]{
0.00, 9.956e-01
0.50, 9.775e-01
1.00, 9.104e-01
1.50, 7.356e-01
2.00, 4.795e-01
2.50, 2.078e-01
3.00, 6.047e-02
3.50, 1.099e-02
4.00, 1.266e-03
4.50, 7.994e-05
};
\label{plot:float_bler}

\addplot[color=magenta,line width = 1pt, dashed,mark=square,mark size=2.5pt, mark options={solid}]
table[col sep=comma]{
0.00, 3.007e-01
0.50, 2.824e-01
1.00, 2.480e-01
1.50, 1.950e-01
2.00, 1.223e-01
2.50, 5.533e-02
3.00, 1.609e-02
3.50, 3.211e-03
4.00, 3.041e-04
4.50, 2.434e-05
};
\label{plot:3bit_ber}

\addplot[color=apfelgruen,line width = 1pt, dashed,mark=x,mark size=2.5pt, mark options={solid}]
table[col sep=comma]{
0.00, 3.093e-01
0.50, 2.941e-01
1.00, 2.733e-01
1.50, 2.389e-01
2.00, 1.684e-01
2.50, 8.495e-02
3.00, 2.728e-02
3.50, 4.652e-03
4.00, 5.233e-04
4.50, 2.390e-05
};
\label{plot:surr_ber}

\addplot[color=black,line width = 1pt, dashed,mark size=2.5pt, mark options={solid}]
table[col sep=comma]{
0.00, 2.644e-01
0.50, 2.348e-01
1.00, 1.890e-01
1.50, 1.290e-01
2.00, 6.558e-02
2.50, 2.195e-02
3.00, 5.441e-03
3.50, 7.668e-04
4.00, 7.589e-05
4.50, 4.357e-06
};
\label{plot:float_ber}

\coordinate (legend) at (axis description cs:0.02,.02);

\end{axis}

\matrix [
fill=white,
fill opacity=0,
text opacity=1,
matrix of nodes,
align =left,
row sep = -1,
column sep = -5,
inner sep= 2,
anchor=south west,
font=\footnotesize,
column 3/.style={anchor=base west},
] at (legend) {
	BLER & BER  &  \\
	\ref{plot:3bit_bler} & \ref{plot:3bit_ber} & 3-bit uniform \\ 
	\ref{plot:surr_bler} & \ref{plot:surr_ber} & 3-bit surrogate \\ 
	\ref{plot:float_bler} & \ref{plot:float_ber} & float \\ 
};

\end{tikzpicture}

%% file: tikz/h_matrices.tex
\begin{tikzpicture}

\node[left] at (0, 1.5in)   (i1) {\LARGE $i = 1$};
\node[left] at (0, -.5in)   (i5) {\LARGE $i = 6$};
\node[left] at (0, -2.5in)   (i10) {\LARGE $i = 10$};

\pgfplotsset{colormap/jet}

\begin{axis}[%
at={(0,0)},
axis equal,
scale only axis,
axis x line=none,
axis y line=none,
xmin=-1,
xmax=308,
y dir=reverse,
ymin=-1,
ymax=176,
title=\Huge $q$,
title style={at={(0.5,0.8)},anchor=south},
]
\addplot [scatter,only marks, mark=*, mark size=1pt,
  point meta=explicit, point meta min=0, point meta max=5]
  table[x=y,y=x,meta=color,col sep=comma]{tikz/q_i0.csv};
\foreach \x in {0,...,14}
{
\edef\temp{\noexpand\draw [black] (axis cs:\x*22-.5,-0.5) -- (axis cs:\x*22-.5,176.5);}
    \temp
}
\foreach \y in {0,...,8}
{
\edef\temp{\noexpand\draw [black] (axis cs:-0.5,\y*22-0.5) -- (axis cs:308.5,\y*22-0.5);}
    \temp
}
\end{axis}

\begin{axis}[%
at={(3.5in,0)},
axis equal,
scale only axis,
axis x line=none,
axis y line=none,
xmin=-1,
xmax=308,
y dir=reverse,
ymin=-1,
ymax=176,
title style={at={(0.5,0.8)},anchor=south},
title=\Huge $r$,
]
\addplot [scatter,only marks, mark=*, mark size=1pt,
  point meta=explicit, point meta min=0, point meta max=5]
  table[x=y,y=x,meta=color,col sep=comma]{tikz/r_i0.csv};
\foreach \x in {0,...,14}
{
\edef\temp{\noexpand\draw [black] (axis cs:\x*22-.5,-0.5) -- (axis cs:\x*22-.5,176.5);}
    \temp
}
\foreach \y in {0,...,8}
{
\edef\temp{\noexpand\draw [black] (axis cs:-0.5,\y*22-0.5) -- (axis cs:308.5,\y*22-0.5);}
    \temp
}
\end{axis}

\begin{axis}[%
at={(0,-2in)},
axis equal,
scale only axis,
axis x line=none,
axis y line=none,
xmin=-1,
xmax=308,
y dir=reverse,
ymin=-1,
ymax=176,
]
\addplot [scatter,only marks, mark=*, mark size=1pt,
  point meta=explicit, point meta min=0, point meta max=5]
  table[x=y,y=x,meta=color,col sep=comma]{tikz/q_i5.csv};
\foreach \x in {0,...,14}
{
\edef\temp{\noexpand\draw [black] (axis cs:\x*22-.5,-0.5) -- (axis cs:\x*22-.5,176.5);}
    \temp
}
\foreach \y in {0,...,8}
{
\edef\temp{\noexpand\draw [black] (axis cs:-0.5,\y*22-0.5) -- (axis cs:308.5,\y*22-0.5);}
    \temp
}
\end{axis}

\begin{axis}[%
at={(3.5in,-2in)},
axis equal,
scale only axis,
axis x line=none,
axis y line=none,
xmin=-1,
xmax=308,
y dir=reverse,
ymin=-1,
ymax=176,
colorbar,
colorbar style={
    height=5.7in,
    at={(1.02,0.48)},
    anchor=west,
}
]
\addplot [scatter,only marks, mark=*, mark size=1pt,
  point meta=explicit, point meta min=0, point meta max=5]
  table[x=y,y=x,meta=color,col sep=comma]{tikz/r_i5.csv};
\foreach \x in {0,...,14}
{
\edef\temp{\noexpand\draw [black] (axis cs:\x*22-.5,-0.5) -- (axis cs:\x*22-.5,176.5);}
    \temp
}
\foreach \y in {0,...,8}
{
\edef\temp{\noexpand\draw [black] (axis cs:-0.5,\y*22-0.5) -- (axis cs:308.5,\y*22-0.5);}
    \temp
}
\end{axis}

\begin{axis}[%
at={(0,-4in)},
axis equal,
scale only axis,
axis x line=none,
axis y line=none,
xmin=-1,
xmax=308,
y dir=reverse,
ymin=-1,
ymax=176,
]
\addplot [scatter,only marks, mark=*, mark size=1pt,
  point meta=explicit, point meta min=0, point meta max=5]
  table[x=y,y=x,meta=color,col sep=comma]{tikz/q_i9.csv};
\foreach \x in {0,...,14}
{
\edef\temp{\noexpand\draw [black] (axis cs:\x*22-.5,-0.5) -- (axis cs:\x*22-.5,176.5);}
    \temp
}
\foreach \y in {0,...,8}
{
\edef\temp{\noexpand\draw [black] (axis cs:-0.5,\y*22-0.5) -- (axis cs:308.5,\y*22-0.5);}
    \temp
}
\end{axis}

\begin{axis}[%
at={(3.5in,-4in)},
axis equal,
scale only axis,
axis x line=none,
axis y line=none,
xmin=-1,
xmax=308,
y dir=reverse,
ymin=-1,
ymax=176,
]
\addplot [scatter,only marks, mark=*, mark size=1pt,
  point meta=explicit, point meta min=0, point meta max=5]
  table[x=y,y=x,meta=color,col sep=comma]{tikz/r_i9.csv};
\foreach \x in {0,...,14}
{
\edef\temp{\noexpand\draw [black] (axis cs:\x*22-.5,-0.5) -- (axis cs:\x*22-.5,176.5);}
    \temp
}
\foreach \y in {0,...,8}
{
\edef\temp{\noexpand\draw [black] (axis cs:-0.5,\y*22-0.5) -- (axis cs:308.5,\y*22-0.5);}
    \temp
}
\end{axis}

\end{tikzpicture}%

%% file: tikz/loss.tex
\begin{tikzpicture}
\begin{axis}[
width=\linewidth,
height=.6\linewidth,
grid style={dotted,anthrazit},
xmajorgrids,
yminorticks=true,
ymajorgrids,
legend columns=1,
legend pos=north east,   
legend cell align={left},
legend style={fill,fill opacity=0, text opacity=1,draw=none},
xlabel={Epoch},
ylabel={Loss $\mathcal{L}$},
legend image post style={mark indices={}},
mark size=1.5pt,
xmin=0,
xmax=3000,
ymin=0.58,
ymax=.7,
]


\addplot[color=mittelblau,line width = 1pt, solid,mark size=2.5pt,mark=x,mark repeat=10, mark options={solid}]
table[col sep=comma]{
0, 6.977e-01
50, 6.997e-01
100, 6.872e-01
150, 6.786e-01
200, 6.701e-01
250, 6.641e-01
300, 6.588e-01
350, 6.532e-01
400, 6.510e-01
450, 6.457e-01
500, 6.448e-01
550, 6.402e-01
600, 6.396e-01
650, 6.367e-01
700, 6.345e-01
750, 6.302e-01
800, 6.291e-01
850, 6.272e-01
900, 6.288e-01
950, 6.250e-01
1000, 6.233e-01
1050, 6.231e-01
1100, 6.224e-01
1150, 6.206e-01
1200, 6.200e-01
1250, 6.172e-01
1300, 6.157e-01
1350, 6.163e-01
1400, 6.149e-01
1450, 6.126e-01
1500, 6.119e-01
1550, 6.111e-01
1600, 6.104e-01
1650, 6.097e-01
1700, 6.104e-01
1750, 6.084e-01
1800, 6.085e-01
1850, 6.075e-01
1900, 6.041e-01
1950, 6.042e-01
2000, 6.042e-01
2050, 6.024e-01
2100, 6.026e-01
2150, 6.018e-01
2200, 6.018e-01
2250, 6.015e-01
2300, 6.000e-01
2350, 6.000e-01
2400, 5.973e-01
2450, 5.978e-01
2500, 5.994e-01
2550, 5.973e-01
2600, 5.970e-01
2650, 5.966e-01
2700, 5.962e-01
2750, 5.956e-01
2800, 5.947e-01
2850, 5.944e-01
2900, 5.931e-01
2950, 5.922e-01
3000, 5.923e-01
};
\label{plot:bgws}
\addlegendentry{\footnotesize No weight sharing (WS)};

\addplot[color=red,line width = 1pt, solid,mark size=2.5pt,mark=diamond,mark repeat=10, mark options={solid}]
table[col sep=comma]{
0, 7.056e-01
50, 6.957e-01
100, 6.780e-01
150, 6.651e-01
200, 6.560e-01
250, 6.497e-01
300, 6.448e-01
350, 6.398e-01
400, 6.367e-01
450, 6.336e-01
500, 6.293e-01
550, 6.282e-01
600, 6.256e-01
650, 6.242e-01
700, 6.227e-01
750, 6.201e-01
800, 6.194e-01
850, 6.177e-01
900, 6.161e-01
950, 6.157e-01
1000, 6.115e-01
1050, 6.120e-01
1100, 6.120e-01
1150, 6.088e-01
1200, 6.075e-01
1250, 6.073e-01
1300, 6.091e-01
1350, 6.044e-01
1400, 6.042e-01
1450, 6.040e-01
1500, 6.035e-01
1550, 6.031e-01
1600, 6.025e-01
1650, 6.023e-01
1700, 6.007e-01
1750, 6.011e-01
1800, 6.009e-01
1850, 5.985e-01
1900, 5.978e-01
1950, 5.989e-01
2000, 5.970e-01
2050, 5.954e-01
2100, 5.961e-01
2150, 5.954e-01
2200, 5.936e-01
2250, 5.948e-01
2300, 5.939e-01
2350, 5.943e-01
2400, 5.926e-01
2450, 5.914e-01
2500, 5.928e-01
2550, 5.907e-01
2600, 5.890e-01
2650, 5.907e-01
2700, 5.906e-01
2750, 5.879e-01
2800, 5.895e-01
2850, 5.887e-01
2900, 5.875e-01
2950, 5.885e-01
3000, 5.877e-01
};
\label{plot:bgws}
\addlegendentry{\footnotesize Base graph weight sharing (BG-WS)};

\addplot[color=apfelgruen,line width = 1pt, solid,mark=o,mark repeat=10,mark size=2.5pt, mark options={solid}]
table[col sep=comma]{
0, 7.086e-01
50, 6.938e-01
100, 6.755e-01
150, 6.612e-01
200, 6.548e-01
250, 6.508e-01
300, 6.465e-01
350, 6.443e-01
400, 6.435e-01
450, 6.409e-01
500, 6.388e-01
550, 6.408e-01
600, 6.389e-01
650, 6.370e-01
700, 6.361e-01
750, 6.387e-01
800, 6.369e-01
850, 6.362e-01
900, 6.370e-01
950, 6.359e-01
1000, 6.345e-01
1050, 6.345e-01
1100, 6.335e-01
1150, 6.318e-01
1200, 6.343e-01
1250, 6.329e-01
1300, 6.333e-01
1350, 6.348e-01
1400, 6.342e-01
1450, 6.323e-01
1500, 6.327e-01
1550, 6.313e-01
1600, 6.324e-01
1650, 6.314e-01
1700, 6.297e-01
1750, 6.307e-01
1800, 6.312e-01
1850, 6.309e-01
1900, 6.319e-01
1950, 6.301e-01
2000, 6.307e-01
2050, 6.298e-01
2100, 6.284e-01
2150, 6.293e-01
2200, 6.300e-01
2250, 6.279e-01
2300, 6.280e-01
2350, 6.295e-01
2400, 6.271e-01
2450, 6.288e-01
2500, 6.279e-01
2550, 6.286e-01
2600, 6.265e-01
2650, 6.268e-01
2700, 6.279e-01
2750, 6.281e-01
2800, 6.276e-01
2850, 6.261e-01
2900, 6.273e-01
2950, 6.283e-01
3000, 6.271e-01
};
\label{plot:layerws}
\addlegendentry{\footnotesize Check node weight sharing (CN-WS)};

\end{axis}

\end{tikzpicture}

%% file: tikz/results_ws_er.tex
\begin{tikzpicture}
\begin{axis}[
width=\linewidth,
height=.8\linewidth,
grid style={dotted,anthrazit},
xmajorgrids,
yminorticks=true,
ymajorgrids,
legend columns=1,
legend pos=south west,   
legend cell align={left},
legend style={fill,fill opacity=0.8},
xlabel={$E_\mathrm{b}/N_0$ [dB]},
ylabel={BER / BLER},
legend image post style={mark indices={}},
ymode=log,
mark size=1.5pt,
xmin=1,
xmax=4.5,
ymin=7e-5,
ymax=1,
]

\addplot[color=magenta,line width = 1pt, solid,mark=square,mark size=2.5pt, mark options={solid}]
table[col sep=comma]{
0.00, 9.993e-01
0.50, 9.973e-01
1.00, 9.719e-01
1.50, 8.860e-01
2.00, 6.741e-01
2.50, 3.811e-01
3.00, 1.438e-01
3.50, 3.278e-02
4.00, 4.344e-03
4.50, 3.888e-04
};
\label{plot:3bit_bler}

\addplot[color=mittelblau,line width = 1pt, solid,mark=x,mark size=2.5pt, mark options={solid}]
table[col sep=comma]{
0.00, 9.998e-01
0.50, 9.976e-01
1.00, 9.751e-01
1.50, 8.831e-01
2.00, 6.692e-01
2.50, 3.474e-01
3.00, 1.128e-01
3.50, 2.409e-02
4.00, 2.574e-03
4.50, 1.990e-04
};
\label{plot:nows_surr_bler}

\addplot[color=rot,line width = 1pt, solid,mark=diamond,mark size=2.5pt, mark options={solid}]
table[col sep=comma]{
0.00, 1.000e+00
0.50, 9.973e-01
1.00, 9.814e-01
1.50, 9.026e-01
2.00, 6.870e-01
2.50, 3.743e-01
3.00, 1.316e-01
3.50, 2.694e-02
4.00, 2.707e-03
4.50, 1.838e-04
};
\label{plot:bg_surr_bler}

\addplot[color=apfelgruen,line width = 1pt, solid,mark=o,mark size=2.5pt, mark options={solid}]
table[col sep=comma]{
0.00, 9.998e-01
0.50, 9.983e-01
1.00, 9.795e-01
1.50, 9.070e-01
2.00, 6.990e-01
2.50, 3.716e-01
3.00, 1.260e-01
3.50, 2.417e-02
4.00, 2.906e-03
4.50, 1.855e-04
};
\label{plot:cn_surr_bler}

\addplot[color=black,line width = 1pt, solid,mark size=2.5pt, mark options={solid}]
table[col sep=comma]{
0.00, 9.956e-01
0.50, 9.775e-01
1.00, 9.104e-01
1.50, 7.356e-01
2.00, 4.795e-01
2.50, 2.078e-01
3.00, 6.047e-02
3.50, 1.099e-02
4.00, 1.266e-03
4.50, 7.994e-05
};
\label{plot:float_bler}

\addplot[color=magenta,line width = 1pt, dashed,mark=square,mark size=2.5pt, mark options={solid}]
table[col sep=comma]{
0.00, 3.007e-01
0.50, 2.824e-01
1.00, 2.480e-01
1.50, 1.950e-01
2.00, 1.223e-01
2.50, 5.533e-02
3.00, 1.609e-02
3.50, 3.211e-03
4.00, 3.041e-04
4.50, 2.434e-05
};
\label{plot:3bit_ber}

\addplot[color=mittelblau,line width = 1pt, dashed,mark=x,mark size=2.5pt, mark options={solid}]
table[col sep=comma]{
0.00, 2.968e-01
0.50, 2.764e-01
1.00, 2.425e-01
1.50, 1.842e-01
2.00, 1.114e-01
2.50, 4.398e-02
3.00, 1.128e-02
3.50, 1.740e-03
4.00, 1.805e-04
4.50, 1.128e-05
};
\label{plot:nows_surr_ber}

\addplot[color=rot,line width = 1pt, dashed,mark=diamnond,mark size=2.5pt, mark options={solid}]
table[col sep=comma]{
0.00, 2.978e-01
0.50, 2.761e-01
1.00, 2.460e-01
1.50, 1.934e-01
2.00, 1.145e-01
2.50, 4.824e-02
3.00, 1.332e-02
3.50, 2.089e-03
4.00, 1.759e-04
4.50, 8.596e-06
};
\label{plot:bg_surr_ber}

\addplot[color=apfelgruen,line width = 1pt, dashed,mark=o,mark size=2.5pt, mark options={solid}]
table[col sep=comma]{
0.00, 2.992e-01
0.50, 2.810e-01
1.00, 2.492e-01
1.50, 1.952e-01
2.00, 1.197e-01
2.50, 4.860e-02
3.00, 1.369e-02
3.50, 2.089e-03
4.00, 1.890e-04
4.50, 9.636e-06
};
\label{plot:cn_surr_ber}

\addplot[color=black,line width = 1pt, dashed,mark size=2.5pt, mark options={solid}]
table[col sep=comma]{
0.00, 2.644e-01
0.50, 2.348e-01
1.00, 1.890e-01
1.50, 1.290e-01
2.00, 6.558e-02
2.50, 2.195e-02
3.00, 5.441e-03
3.50, 7.668e-04
4.00, 7.589e-05
4.50, 4.357e-06
};
\label{plot:float_ber}

\coordinate (legend) at (axis description cs:0.02,.02);

\end{axis}

\matrix [
fill=white,
fill opacity=0,
text opacity=1,
matrix of nodes,
align =left,
row sep = -1,
column sep = -5,
inner sep= 2,
anchor=south west,
font=\footnotesize,
column 3/.style={anchor=base west},
] at (legend) {
	BLER & BER  &  \\
	\ref{plot:3bit_bler} & \ref{plot:3bit_ber} & 3-bit uniform \\ 
	\ref{plot:nows_surr_bler} & \ref{plot:nows_surr_ber} & trained surrogate, no WS \\ 
	\ref{plot:bg_surr_bler} & \ref{plot:bg_surr_ber} & trained surrogate, BG-WS \\ 
	\ref{plot:cn_surr_bler} & \ref{plot:cn_surr_ber} & trained surrogate, CN-WS \\ 
	\ref{plot:float_bler} & \ref{plot:float_ber} & float \\ 
};

\end{tikzpicture}

%% file: tikz/results_ws_er_conv.tex
\begin{tikzpicture}
\begin{axis}[
width=\linewidth,
height=.8\linewidth,
grid style={dotted,anthrazit},
xmajorgrids,
yminorticks=true,
ymajorgrids,
legend columns=1,
legend pos=south west,   
legend cell align={left},
legend style={fill,fill opacity=0.8},
xlabel={$E_\mathrm{b}/N_0$ [dB]},
ylabel={BER / BLER},
legend image post style={mark indices={}},
ymode=log,
mark size=1.5pt,
xmin=1,
xmax=4,
ymin=7e-5,
ymax=1,
]

\addplot[color=magenta,line width = 1pt, solid,mark=square,mark size=2.5pt, mark options={solid}]
table[col sep=comma]{
0.00, 9.993e-01
0.50, 9.973e-01
1.00, 9.719e-01
1.50, 8.860e-01
2.00, 6.741e-01
2.50, 3.811e-01
3.00, 1.438e-01
3.50, 3.278e-02
4.00, 4.344e-03
};
\label{plot:3bit_bler}

\addplot[color=mittelblau,line width = 1pt, solid,mark=x,mark size=2.5pt, mark options={solid}]
table[col sep=comma]{
0.00, 9.988e-01
0.50, 9.890e-01
1.00, 9.429e-01
1.50, 7.966e-01
2.00, 5.479e-01
2.50, 2.886e-01
3.00, 9.717e-02
3.50, 2.181e-02
4.00, 5.837e-03
};
\label{plot:nows_conv_bler}

\addplot[color=rot,line width = 1pt, solid,mark=diamond,mark size=2.5pt, mark options={solid}]
table[col sep=comma]{
0.00, 9.993e-01
0.50, 9.915e-01
1.00, 9.448e-01
1.50, 8.132e-01
2.00, 5.627e-01
2.50, 2.817e-01
3.00, 9.229e-02
3.50, 2.376e-02
4.00, 5.815e-03
};
\label{plot:bg_conv_bler}

\addplot[color=apfelgruen,line width = 1pt, solid,mark=o,mark size=2.5pt, mark options={solid}]
table[col sep=comma]{
0.00, 9.993e-01
0.50, 9.941e-01
1.00, 9.539e-01
1.50, 8.181e-01
2.00, 5.725e-01
2.50, 2.842e-01
3.00, 9.058e-02
3.50, 2.140e-02
4.00, 3.277e-03
};
\label{plot:cn_conv_bler}

\addplot[color=black,line width = 1pt, solid,mark size=2.5pt, mark options={solid}]
table[col sep=comma]{
0.00, 9.956e-01
0.50, 9.775e-01
1.00, 9.104e-01
1.50, 7.356e-01
2.00, 4.795e-01
2.50, 2.078e-01
3.00, 6.047e-02
3.50, 1.099e-02
4.00, 1.266e-03
};
\label{plot:float_bler}

\addplot[color=magenta,line width = 1pt, dashed,mark=square,mark size=2.5pt, mark options={solid}]
table[col sep=comma]{
0.00, 3.007e-01
0.50, 2.824e-01
1.00, 2.480e-01
1.50, 1.950e-01
2.00, 1.223e-01
2.50, 5.533e-02
3.00, 1.609e-02
3.50, 3.211e-03
4.00, 3.041e-04
};
\label{plot:3bit_ber}

\addplot[color=mittelblau,line width = 1pt, dashed,mark=x,mark size=2.5pt, mark options={solid}]
table[col sep=comma]{
0.00, 2.439e-01
0.50, 2.209e-01
1.00, 1.843e-01
1.50, 1.282e-01
2.00, 6.807e-02
2.50, 2.846e-02
3.00, 6.819e-03
3.50, 9.624e-04
4.00, 1.686e-04
};
\label{plot:nows_conv_ber}

\addplot[color=rot,line width = 1pt, dashed,mark=diamond,mark size=2.5pt, mark options={solid}]
table[col sep=comma]{
0.00, 2.401e-01
0.50, 2.160e-01
1.00, 1.794e-01
1.50, 1.255e-01
2.00, 6.955e-02
2.50, 2.628e-02
3.00, 6.938e-03
3.50, 9.655e-04
4.00, 1.572e-04
};
\label{plot:bg_conv_ber};

\addplot[color=apfelgruen,line width = 1pt, dashed,mark=o,mark size=2.5pt, mark options={solid}]
table[col sep=comma]{
0.00, 2.578e-01
0.50, 2.322e-01
1.00, 1.923e-01
1.50, 1.368e-01
2.00, 7.491e-02
2.50, 2.957e-02
3.00, 7.315e-03
3.50, 1.305e-03
4.00, 1.548e-04
};
\label{plot:cn_conv_ber}

\addplot[color=black,line width = 1pt, dashed,mark size=2.5pt, mark options={solid}]
table[col sep=comma]{
0.00, 2.644e-01
0.50, 2.348e-01
1.00, 1.890e-01
1.50, 1.290e-01
2.00, 6.558e-02
2.50, 2.195e-02
3.00, 5.441e-03
3.50, 7.668e-04
4.00, 7.589e-05
};
\label{plot:float_ber}

\coordinate (legend) at (axis description cs:0.02,.02);

\end{axis}

\matrix [
fill=white,
fill opacity=0,
text opacity=1,
matrix of nodes,
align =left,
row sep = -1,
column sep = -5,
inner sep= 2,
anchor=south west,
font=\footnotesize,
column 3/.style={anchor=base west},
] at (legend) {
	BLER & BER  &  \\
	\ref{plot:3bit_bler} & \ref{plot:3bit_ber} & 3-bit uniform \\ 
	\ref{plot:nows_conv_bler} & \ref{plot:nows_conv_ber} & trained converted, no WS \\ 
	\ref{plot:bg_conv_bler} & \ref{plot:bg_conv_ber} & trained converted, BG-WS \\ 
	\ref{plot:cn_conv_bler} & \ref{plot:cn_conv_ber} & trained converted, CN-WS \\ 
	\ref{plot:float_bler} & \ref{plot:float_ber} & float \\ 
};

\end{tikzpicture}

%% file: tikz/results_rayleigh.tex
\begin{tikzpicture}
\begin{axis}[
width=\linewidth,
height=.8\linewidth,
grid style={dotted,anthrazit},
xmajorgrids,
yminorticks=true,
ymajorgrids,
legend columns=1,
legend pos=south west,   
legend cell align={left},
legend style={fill,fill opacity=0.8},
xlabel={$E_\mathrm{b}/N_0$ [dB]},
ylabel={BER / BLER},
legend image post style={mark indices={}},
ymode=log,
mark size=1.5pt,
xmin=2,
xmax=7,
ymin=3e-4,
ymax=1,
]

\addplot[color=magenta,line width = 1pt, solid,mark=square,mark size=2.5pt, mark options={solid}]
table[col sep=comma]{
0.00, 1.000e+00
0.50, 1.000e+00
1.00, 9.998e-01
1.50, 9.998e-01
2.00, 9.985e-01
2.50, 9.902e-01
3.00, 9.680e-01
3.50, 9.111e-01
4.00, 7.925e-01
4.50, 6.199e-01
5.00, 4.243e-01
5.50, 2.432e-01
6.00, 1.201e-01
6.50, 4.744e-02
7.00, 1.703e-02
};
\label{plot:r3bit_bler}

\addplot[color=apfelgruen,line width = 1pt, solid,mark=o,mark size=2.5pt, mark options={solid}]
table[col sep=comma]{
0.00, 1.000e+00
0.50, 1.000e+00
1.00, 1.000e+00
1.50, 9.995e-01
2.00, 9.971e-01
2.50, 9.902e-01
3.00, 9.629e-01
3.50, 8.857e-01
4.00, 7.515e-01
4.50, 5.596e-01
5.00, 3.398e-01
5.50, 1.719e-01
6.00, 8.496e-02
6.50, 2.959e-02
7.00, 8.963e-03
};
\label{plot:rconv_bler}

\addplot[color=mittelblau,line width = 1pt, solid,mark=x,mark size=2.5pt, mark options={solid}]
table[col sep=comma]{
0.00, 1.000e+00
0.50, 1.000e+00
1.00, 1.000e+00
1.50, 1.000e+00
2.00, 9.990e-01
2.50, 9.917e-01
3.00, 9.580e-01
3.50, 8.884e-01
4.00, 7.651e-01
4.50, 5.710e-01
5.00, 3.481e-01
5.50, 1.865e-01
6.00, 7.031e-02
6.50, 2.791e-02
7.00, 9.417e-03
};
\label{plot:rconv_g_bler}

\addplot[color=black,line width = 1pt, solid,mark size=2.5pt, mark options={solid}]
table[col sep=comma]{
0.00, 1.000e+00
0.50, 1.000e+00
1.00, 1.000e+00
1.50, 9.973e-01
2.00, 9.875e-01
2.50, 9.685e-01
3.00, 9.121e-01
3.50, 8.013e-01
4.00, 6.152e-01
4.50, 4.141e-01
5.00, 2.373e-01
5.50, 1.082e-01
6.00, 4.305e-02
6.50, 1.367e-02
7.00, 3.495e-03
};
\label{plot:rfloat_bler}

\addplot[color=magenta,line width = 1pt, dashed,mark=square,mark size=2.5pt, mark options={solid}]
table[col sep=comma]{
0.00, 3.229e-01
0.50, 3.155e-01
1.00, 3.051e-01
1.50, 2.944e-01
2.00, 2.830e-01
2.50, 2.638e-01
3.00, 2.366e-01
3.50, 2.019e-01
4.00, 1.526e-01
4.50, 1.025e-01
5.00, 5.791e-02
5.50, 2.855e-02
6.00, 1.235e-02
6.50, 4.507e-03
7.00, 1.406e-03
};
\label{plot:r3bit_ber}

\addplot[color=apfelgruen,line width = 1pt, dashed,mark=o,mark size=2.5pt, mark options={solid}]
table[col sep=comma]{
0.00, 2.981e-01
0.50, 2.857e-01
1.00, 2.701e-01
1.50, 2.545e-01
2.00, 2.409e-01
2.50, 2.186e-01
3.00, 1.908e-01
3.50, 1.548e-01
4.00, 1.136e-01
4.50, 7.306e-02
5.00, 3.664e-02
5.50, 1.456e-02
6.00, 6.268e-03
6.50, 2.082e-03
7.00, 4.875e-04
};
\label{plot:rconv_ber}

\addplot[color=mittelblau,line width = 1pt, dashed,mark=x,mark size=2.5pt, mark options={solid}]
table[col sep=comma]{
0.00, 2.969e-01
0.50, 2.830e-01
1.00, 2.674e-01
1.50, 2.541e-01
2.00, 2.370e-01
2.50, 2.151e-01
3.00, 1.884e-01
3.50, 1.539e-01
4.00, 1.134e-01
4.50, 7.191e-02
5.00, 3.815e-02
5.50, 1.744e-02
6.00, 5.456e-03
6.50, 2.036e-03
7.00, 5.409e-04
};
\label{plot:rconv_g_ber}

\addplot[color=black,line width = 1pt, dashed,mark size=2.5pt, mark options={solid}]
table[col sep=comma]{
0.00, 3.090e-01
0.50, 2.982e-01
1.00, 2.872e-01
1.50, 2.717e-01
2.00, 2.522e-01
2.50, 2.224e-01
3.00, 1.834e-01
3.50, 1.384e-01
4.00, 9.180e-02
4.50, 5.276e-02
5.00, 2.531e-02
5.50, 1.015e-02
6.00, 3.563e-03
6.50, 1.002e-03
7.00, 2.350e-04
};
\label{plot:rfloat_ber}

\coordinate (legend) at (axis description cs:0.02,.02);

\end{axis}

\matrix [
fill=white,
fill opacity=0,
text opacity=1,
matrix of nodes,
align =left,
row sep = -1,
column sep = -5,
inner sep= 2,
anchor=south west,
font=\footnotesize,
column 3/.style={anchor=base west},
] at (legend) {
	BLER & BER  &  \\
	\ref{plot:r3bit_bler} & \ref{plot:r3bit_ber} & 3-bit uniform \\ 
	\ref{plot:rconv_bler} & \ref{plot:rconv_ber} & converted, trained @ 5 dB Rayleigh \\ 
	\ref{plot:rconv_g_bler} & \ref{plot:rconv_g_ber} & converted, generalized from AWGN \\ 
    \ref{plot:rfloat_bler} & \ref{plot:rfloat_ber} & float \\ 
};

\end{tikzpicture}

%% file: tikz/results_1_4.tex
\begin{tikzpicture}
\begin{axis}[
width=\linewidth,
height=.8\linewidth,
grid style={dotted,anthrazit},
xmajorgrids,
yminorticks=true,
ymajorgrids,
legend columns=1,
legend pos=south west,   
legend cell align={left},
legend style={fill,fill opacity=0.8},
xlabel={$E_\mathrm{b}/N_0$ [dB]},
ylabel={BER / BLER},
legend image post style={mark indices={}},
ymode=log,
mark size=1.5pt,
xmin=1,
xmax=4,
ymin=1e-5,
ymax=1,
]

\addplot[color=magenta,line width = 1pt, solid,mark=square,mark size=2.5pt, mark options={solid}]
table[col sep=comma]{
0.00, 9.998e-01
0.50, 9.939e-01
1.00, 9.575e-01
1.50, 8.232e-01
2.00, 5.386e-01
2.50, 2.261e-01
3.00, 4.997e-02
3.50, 6.274e-03
4.00, 4.854e-04
};
\label{plot:3bit_1_4_bler}

\addplot[color=apfelgruen,line width = 1pt, solid,mark=o,mark size=2.5pt, mark options={solid}]
table[col sep=comma]{
0.00, 9.976e-01
0.50, 9.824e-01
1.00, 8.948e-01
1.50, 6.646e-01
2.00, 3.425e-01
2.50, 9.814e-02
3.00, 1.660e-02
3.50, 1.532e-03
4.00, 6.286e-05
};
\label{plot:surr_1_4_bler}

\addplot[color=mittelblau,line width = 1pt, solid,mark=x,mark size=2.5pt, mark options={solid}]
table[col sep=comma]{
0.00, 9.990e-01
0.50, 9.883e-01
1.00, 9.072e-01
1.50, 6.934e-01
2.00, 3.936e-01
2.50, 1.179e-01
3.00, 2.262e-02
3.50, 1.726e-03
4.00, 1.023e-04
};
\label{plot:conv_1_4_bler}

\addplot[color=black,line width = 1pt, solid,mark size=2.5pt, mark options={solid}]
table[col sep=comma]{
0.00, 9.829e-01
0.50, 9.336e-01
1.00, 7.725e-01
1.50, 4.788e-01
2.00, 2.034e-01
2.50, 5.306e-02
3.00, 7.229e-03
3.50, 5.651e-04
4.00, 3.247e-05
};
\label{plot:float_1_4_bler}

\addplot[color=magenta,line width = 1pt, dashed,mark=square,mark size=2.5pt, mark options={solid}]
table[col sep=comma]{
0.00, 3.492e-01
0.50, 3.174e-01
1.00, 2.593e-01
1.50, 1.719e-01
2.00, 8.070e-02
2.50, 2.404e-02
3.00, 3.372e-03
3.50, 3.566e-04
4.00, 1.957e-05
};
\label{plot:3bit_1_4_ber}

\addplot[color=apfelgruen,line width = 1pt, dashed,mark=o,mark size=2.5pt, mark options={solid}]
table[col sep=comma]{
0.00, 3.069e-01
0.50, 2.583e-01
1.00, 1.858e-01
1.50, 9.690e-02
2.00, 3.519e-02
2.50, 6.904e-03
3.00, 8.513e-04
3.50, 5.683e-05
4.00, 2.242e-06
};
\label{plot:surr_1_4_ber}

\addplot[color=mittelblau,line width = 1pt, dashed,mark=x,mark size=2.5pt, mark options={solid}]
table[col sep=comma]{
0.00, 3.205e-01
0.50, 2.800e-01
1.00, 2.077e-01
1.50, 1.110e-01
2.00, 4.286e-02
2.50, 8.695e-03
3.00, 1.158e-03
3.50, 7.185e-05
4.00, 2.422e-06
};
\label{plot:conv_1_4_ber}

\addplot[color=black,line width = 1pt, dashed,mark size=2.5pt, mark options={solid}]
table[col sep=comma]{
0.00, 2.750e-01
0.50, 2.148e-01
1.00, 1.343e-01
1.50, 5.695e-02
2.00, 1.780e-02
2.50, 3.480e-03
3.00, 3.791e-04
3.50, 2.108e-05
4.00, 9.459e-07
};
\label{plot:float_1_4_ber}

\coordinate (legend) at (axis description cs:0.02,.02);

\end{axis}

\matrix [
fill=white,
fill opacity=0,
text opacity=1,
matrix of nodes,
align =left,
row sep = -1,
column sep = -5,
inner sep= 2,
anchor=south west,
font=\footnotesize,
column 3/.style={anchor=base west},
] at (legend) {
	BLER & BER  &  \\
	\ref{plot:3bit_1_4_bler} & \ref{plot:3bit_1_4_ber} & 3-bit uniform \\ 
	\ref{plot:surr_1_4_bler} & \ref{plot:surr_1_4_ber} & trained surrogate, 3.6 bits \\ 
	\ref{plot:conv_1_4_bler} & \ref{plot:conv_1_4_ber} & trained converted, 3.6 bits \\ 
	\ref{plot:float_1_4_bler} & \ref{plot:float_1_4_ber} & float \\ 
};

\end{tikzpicture}

%% file: tikz/results_2_3.tex
\begin{tikzpicture}
\begin{axis}[
width=\linewidth,
height=.8\linewidth,
grid style={dotted,anthrazit},
xmajorgrids,
yminorticks=true,
ymajorgrids,
legend columns=1,
legend pos=south west,   
legend cell align={left},
legend style={fill,fill opacity=0.8},
xlabel={$E_\mathrm{b}/N_0$ [dB]},
ylabel={BER / BLER},
legend image post style={mark indices={}},
ymode=log,
mark size=1.5pt,
xmin=0,
xmax=4.5,
ymin=3e-5,
ymax=1,
]

\addplot[color=magenta,line width = 1pt, solid,mark=square,mark size=2.5pt, mark options={solid}]
table[col sep=comma]{
0.00, 9.998e-01
0.50, 9.939e-01
1.00, 9.673e-01
1.50, 8.965e-01
2.00, 7.292e-01
2.50, 4.556e-01
3.00, 2.151e-01
3.50, 6.506e-02
4.00, 1.440e-02
4.50, 2.081e-03
};
\label{plot:4bit_2_3_bler}

\addplot[color=apfelgruen,line width = 1pt, solid,mark=o,mark size=2.5pt, mark options={solid}]
table[col sep=comma]{
0.00, 9.995e-01
0.50, 9.949e-01
1.00, 9.714e-01
1.50, 8.943e-01
2.00, 7.114e-01
2.50, 4.590e-01
3.00, 2.061e-01
3.50, 7.178e-02
4.00, 1.270e-02
4.50, 1.702e-03
};
\label{plot:conv_2_3_bler}

\addplot[color=mittelblau,line width = 1pt, solid,mark=x,mark size=2.5pt, mark options={solid}]
table[col sep=comma]{
0.00, 9.990e-01
0.50, 9.956e-01
1.00, 9.805e-01
1.50, 9.316e-01
2.00, 7.725e-01
2.50, 5.161e-01
3.00, 2.437e-01
3.50, 7.397e-02
4.00, 1.703e-02
4.50, 2.270e-03
};
\label{plot:conv_g_2_3_bler}

\addplot[color=black,line width = 1pt, solid,mark size=2.5pt, mark options={solid}]
table[col sep=comma]{
0.00, 9.976e-01
0.50, 9.917e-01
1.00, 9.575e-01
1.50, 8.723e-01
2.00, 6.768e-01
2.50, 4.089e-01
3.00, 1.929e-01
3.50, 5.501e-02
4.00, 1.116e-02
4.50, 1.332e-03
};
\label{plot:float_2_3_bler}

\addplot[color=magenta,line width = 1pt, dashed,mark=square,mark size=2.5pt, mark options={solid}]
table[col sep=comma]{
0.00, 2.381e-01
0.50, 2.164e-01
1.00, 1.822e-01
1.50, 1.426e-01
2.00, 9.142e-02
2.50, 4.334e-02
3.00, 1.698e-02
3.50, 3.866e-03
4.00, 6.157e-04
4.50, 6.580e-05
};
\label{plot:4bit_2_3_ber}

\addplot[color=apfelgruen,line width = 1pt, dashed,mark=o,mark size=2.5pt, mark options={solid}]
table[col sep=comma]{
0.00, 2.311e-01
0.50, 2.086e-01
1.00, 1.814e-01
1.50, 1.377e-01
2.00, 8.786e-02
2.50, 4.403e-02
3.00, 1.498e-02
3.50, 3.967e-03
4.00, 5.290e-04
4.50, 5.281e-05
};
\label{plot:conv_2_3_ber}

\addplot[color=mittelblau,line width = 1pt, dashed,mark=x,mark size=2.5pt, mark options={solid}]
table[col sep=comma]{
0.00, 2.390e-01
0.50, 2.173e-01
1.00, 1.885e-01
1.50, 1.471e-01
2.00, 9.972e-02
2.50, 4.839e-02
3.00, 1.783e-02
3.50, 3.941e-03
4.00, 6.677e-04
4.50, 7.309e-05
};
\label{plot:conv_g_2_3_ber}

\addplot[color=black,line width = 1pt, dashed,mark size=2.5pt, mark options={solid}]
table[col sep=comma]{
0.00, 2.263e-01
0.50, 2.032e-01
1.00, 1.711e-01
1.50, 1.289e-01
2.00, 8.008e-02
2.50, 3.745e-02
3.00, 1.416e-02
3.50, 2.842e-03
4.00, 4.873e-04
4.50, 4.747e-05
};
\label{plot:float_2_3_ber}

\coordinate (legend) at (axis description cs:0.02,.02);

\end{axis}

\matrix [
fill=white,
fill opacity=0,
text opacity=1,
matrix of nodes,
align =left,
row sep = -1,
column sep = -5,
inner sep= 2,
anchor=south west,
font=\footnotesize,
column 3/.style={anchor=base west},
] at (legend) {
	BLER & BER  &  \\
	\ref{plot:4bit_2_3_bler} & \ref{plot:4bit_2_3_ber} & 4-bit uniform \\ 
	\ref{plot:conv_2_3_bler} & \ref{plot:conv_2_3_ber} & converted, trained @ $R'=2/3$ \\ 
	\ref{plot:conv_g_2_3_bler} & \ref{plot:conv_g_2_3_ber} & converted, generalized from $R=1/4$ \\ 
    \ref{plot:float_2_3_bler} & \ref{plot:float_2_3_ber} & float \\ 
};

\end{tikzpicture}